%% file: main.tex
\documentclass[a4paper,11pt]{article}
\usepackage{pos}
\pdfoutput=1
\input{macros.tex}

\title{Four-dimensional domain decomposition for the factorization of the fermion determinant}

\author*{Matteo Saccardi}
\author{Leonardo Giusti}

\affiliation{Dipartimento di Fisica, Universit\`a di Milano--Bicocca,\\
and INFN, sezione di Milano--Bicocca,\\
Piazza della Scienza 3, I-20126 Milano, Italy}

\emailAdd{m.saccardi@campus.unimib.it}
\emailAdd{leonardo.giusti@unimib.it}

\abstract
{
The non-local dependence of the fermion determinant on the gauge field limits our ability of simulating Quantum Chromodynamics on the lattice. Here we present a factorization of the gauge field dependence of the fermion determinant based on an overlapping four-dimensional domain decomposition of the lattice. The resulting action is block-local in the gauge and in the auxiliary bosonic fields. Possible applications are multi-level integration, master field simulations, and more efficient parallelizations of Monte Carlo algorithms and codes.
}

\FullConference{%
  The 39th International Symposium on Lattice Field Theory (Lattice2022),\\
  8-13 August, 2022 \\
  Bonn, Germany 
}

\begin{document}

\maketitle

\section{Introduction and motivations\label{sec:intro}}
\input intro.tex

\section{Four--dimensional decomposition of the lattice\label{sec:2}}
\input sec2.tex

\section{Block--decomposition of the fermion determinant\label{sec:3}}
\input sec3.tex

\section{Multi--boson representation and factorization\label{sec:4}}
\input sec4.tex

\section{Conclusions and outlook}
\input concl.tex




\bibliographystyle{JHEP}
\bibliography{mb.bib}

\end{document}

%% file: macros.tex
\newcommand{\be}{\begin{equation}}
\newcommand{\ee}{\end{equation}}
\newcommand{\bea}{\begin{eqnarray}}
\newcommand{\eea}{\end{eqnarray}}
\newcommand{\bes}{\begin{eqnarray}}
\newcommand{\ees}{\end{eqnarray}}
\newcommand{\ba}{\begin{array}}
\newcommand{\ea}{\end{array}}

\def\diracstar#1#2{
    \setbox0=\hbox{$\gamma$}\setbox1=\hbox{$\gamma_{#1}$}
    \gamma_{#1}\kern-\wd1\kern\wd0
    \smash{\raise4.5pt\hbox{$\scriptstyle#2$}}}

%% file: intro.tex
In order to simulate an exactcly gauge invariant field theory on the lattice, we usually introduce link variables $U_\mu(x)$ that belong to the symmetry group manifold and that live on the links between the two adjacent lattice sites $(x,x+\hat \mu)$. Their action $S_G[U]$ is a local quantity, i.e. only neighboring link variables interact. When we introduce matter fields, the corresponding quark field variables $\psi(x)$, $\bar\psi(x)$ reside on the lattice sites. Their action is quadratic and takes the form
\be
S_F[\psi,\bar\psi,U] = a^4 \sum_x \bar\psi(x) \left( D[U] \psi \right) (x) \, .
\ee
It is possible to define different expressions for the lattice Dirac operator $D$, but here we are only interested in local ones. This manifest locality is spoiled once we analytically integrate out the Grassman variables representing the fermionic degrees of freedom, resulting in a global contribution $S_F^{\text{eff}}[U]$ to the effective gluonic action
\be
	S_G^{\text{eff}}[U] = S_G[U] + S_F^{\text{eff}}[U], \quad S_F^{\text{eff}}[U] = - \ln\det D[U] \, ,
\ee
thus limiting our ability of directly simulating fermions on the lattice. In principle, we would expect the notion of locality not to be completely lost, but rather masked out by the complicated functional $\ln\det D[U]$. Intuitively, we would expect it to be possible to approximate it as the sum of local contributions, with an extra contribution which is expected to be tiny since the effective link interaction decreases with the distance of the links.

In this talk we discuss how our intuition can be formalized by following the same steps that were thoroughly described in Ref.~\cite{GIUSTI2022137103}. In Section~\ref{sec:2} we define a meticolously chosen overlapping four-dimensional domain decomposition of the lattice which allows to derive in Section~\ref{sec:3} a block-decomposition of the fermion determinant. We also show that this result is a (not-so-straightforward) generalization of the factorization based on a one-dimensional overlapping domain decomposition that was proposed a few years ago~\cite{Ce:2016idq,Ce:2016ajy}, see also~\cite{Luscher:2003qa,Luscher:2004pav} and that has already been extensively tested numerically~\cite{Ce:2016idq,Ce:2016ajy,Giusti:2017ksp,Ce:2017ndt,DallaBrida:2020cik}. The remaining global contribution is studied in Section~\ref{sec:4} by means of a multi-boson representation, where the resulting multi-boson action is factorized as well. The small residual global term can be regarded as a reweighting factor to be included in the observable, but in principle we could also drop it out and consider a Metropolis-Hastings accept-reject step in order to correct for this approximation.

Besides a purely theoretical interest for studying the factorization of $\det D$, it could in principle have many possible applications, paving the way to multi-level intergation schemes and facilitating master-field simulations~\cite{Luscher:2017cjh,Giusti:2018cmp,Francis:2019muy}, possibly improving the sampling in flow-based generative models~\cite{Albergo_2019,Kanwar_2020,Abbott_2022}, while also allowing for a better and improved parallelization of our codes and algorithms.

%% file: sec2.tex
\begin{figure}[t!]
\begin{center}
\includegraphics[width=0.8\columnwidth]{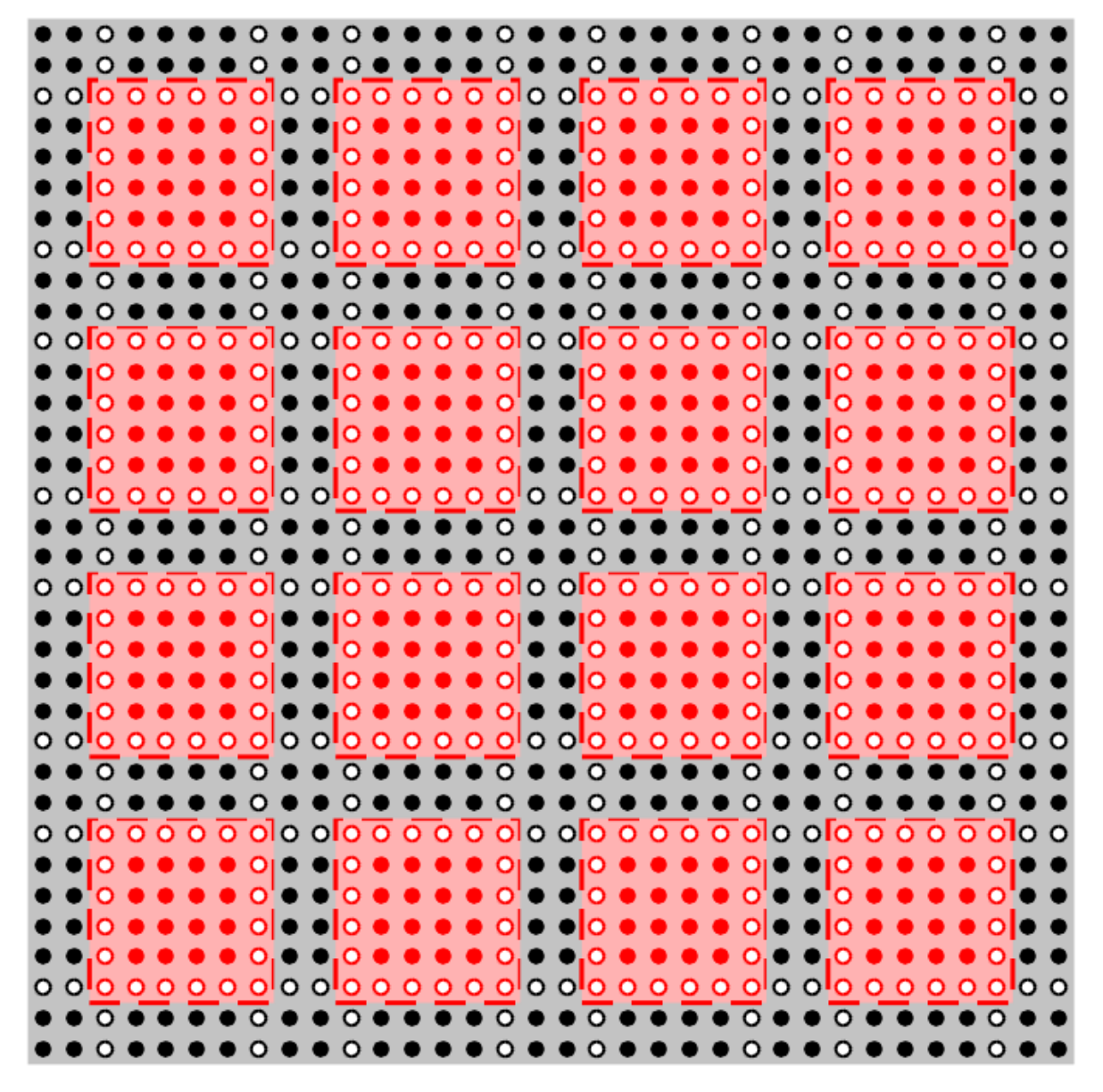}
\caption{Two-dimensional representation of the basic domain decomposition of the lattice in the disconnected domain $\Lambda_{0}$ (red blocks) and the globally connected one $\Lambda_{1}$ (grey thick frame). The empty circles indicate the domain of hyperplanes $\partial \Pi$, with the red and black circles indicating $\partial\Lambda_{0}$ and $\partial \Pi_1$ respectively.
  \label{Fig:startingDD}}
\end{center}
\end{figure}

The factorization of $\det D$ highly depends on the corresponding domain decomposition of the lattice. Our choice is shown in Fig.~\ref{Fig:startingDD} and it is based on three steps.
\begin{enumerate}
\item The first step consists in the division of the lattice in local, hyperrectangular blocks $\Lambda_0^{\hat a}$, where the index $\hat a$ identifies each block. Our final goal is to factorize the dependence of $\det D$ on the gauge-field within these domains, which we collectively define as the global, disconnected ``active'' region
\be
	\Lambda_0 = \bigcup_{\hat a} \Lambda_0^{\hat a} \, .
\ee
The remaining global, connected part of the lattice forms the ``inactive'' domain $\Lambda_1$, which can be interpreted as the connected frame of $\Lambda_0$.

\item The second step starts with the simple observation that the two domains $\Lambda_0$ and $\Lambda_1$ only interact through the internal boundaries $\partial\Lambda_0^{\hat a}$ of each local cell. The global, disconnected domain
\be
\partial\Lambda_0 = \bigcup_{\hat a} \partial\Lambda_0^{\hat a}
\ee
is then formed by the disjoint union of local boundaries $\partial\Lambda_0^{\hat a}$. As it is clear from Fig.~\ref{Fig:startingDD}, the local boundaries lie on orthogonal hyperplanes, the union of which forms the global, connected domain $\partial\Pi$. Different internal boundaries are then connected through the global, disconnected domain $\partial\Pi_1 = \partial\Pi \backslash \partial\Lambda_0$.

\begin{figure}[t!]
\begin{center}
\includegraphics[width=0.33\columnwidth]{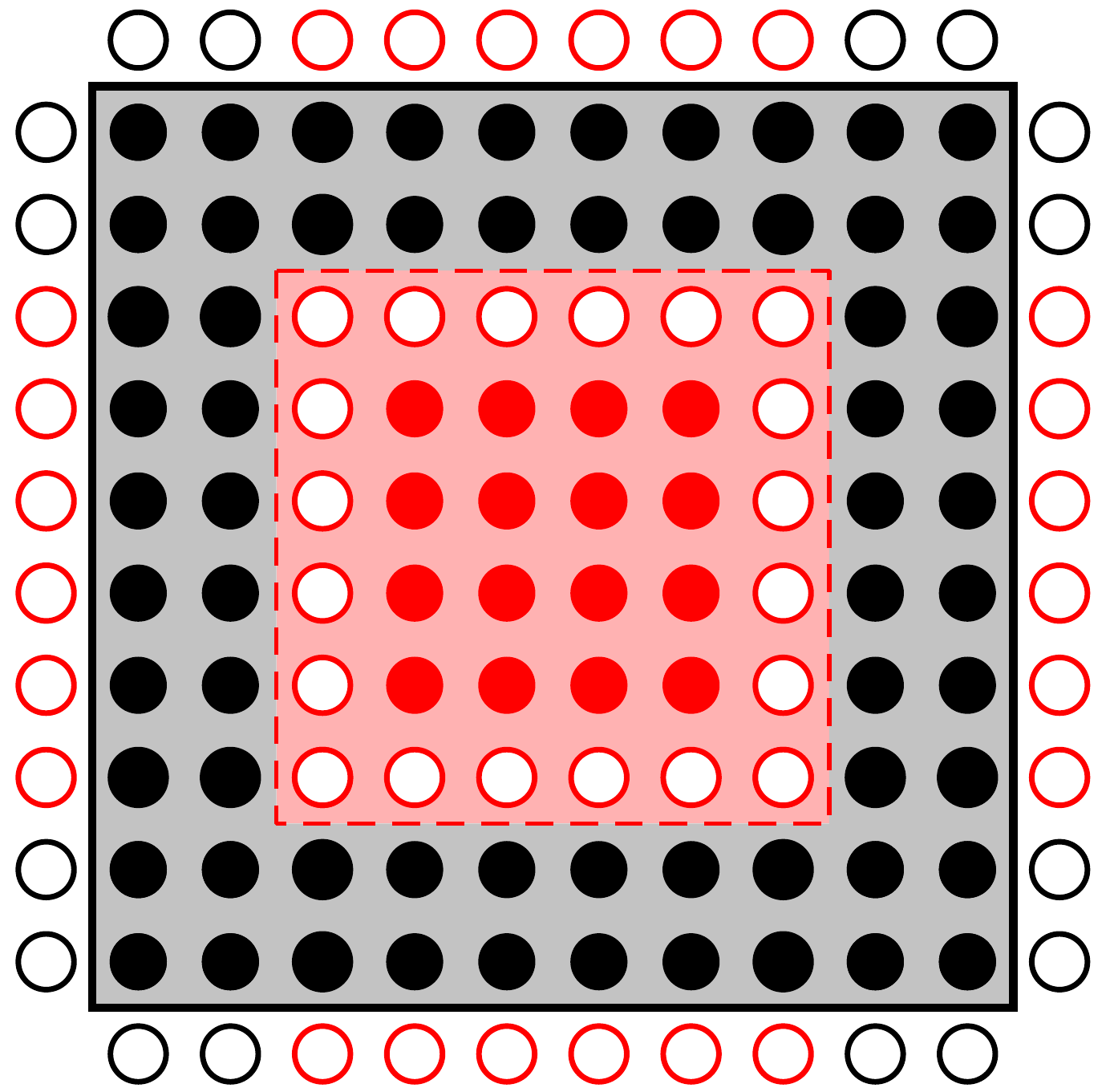}
\caption{A ``framed domain'' $\Omega^{\hat a}_0$ made by the union of a hyperrectangle $\Lambda_{0}^{\hat a}$ (red)
  and its frame $\Phi^{\hat a}_1$ (grey). The points of its exterior boundary $\partial \Omega^{\hat a *}_0$ are
  indicated with open circles outside the continuous black line. These circles are red, if they belong to
  $\partial \Lambda_{0}$, or black , if they belong to  $\partial\Pi_1$ and in particular to
  $\partial \bar\Omega^{\hat a *}_0$.\label{Fig:Omega0a}}
\end{center}
\end{figure}

\item When we combine the definition of local cells in the first step with the definition of the hyperplanes in the second step, we can easily identify a region of the domain $\Lambda_1$ that surrounds each local cell $\Lambda_0^{\hat a}$ and that is confined between these hyperplanes. As shown in Fig.~\ref{Fig:Omega0a}, we define this local frame as $\Phi_1^{\hat a} \in \Lambda_1$ and the corresponding local (framed) cell as
\be
	\Omega_0^{\hat a} = \Lambda_0^{\hat a} \cup \Phi_1^{\hat a} \, .
\ee
The union of these domains forms the anticipated four-dimensional overlapping domain decomposition of the whole lattice. As it is clear from Fig.~\ref{Fig:Omega0a}, if we consider the external boundary of each framed domain, $\partial\bar\Omega_0^{\hat a}$, we can notice that it contains some contributions from $\partial\Lambda_0$ and some from $\partial\Pi_1$, which we identify as $\partial\bar\Omega_0^{\hat a \ast} = \partial\bar\Omega_0^{\hat a} \cap \partial\Pi_1$.
\end{enumerate}
For a more precise definition of these domains we refer to Appendix B of Ref.~\cite{GIUSTI2022137103}.

%% file: sec3.tex
In this section, we want to discuss the block-factorization of $\det D$ that results from the domain decomposition defined in the previous section. We can consider $D$ to be the Wilson--Dirac operator defined in~~\cite{Wilson:1974sk,Sheikholeslami:1985ij}, see also~\cite{Luscher:1996sc} and Appendix A of Ref.~\cite{GIUSTI2022137103}, but in principle we only require it to be local. Here we present a sketch of the derivation that was thoroughly described in Section 3 of Ref.~\cite{GIUSTI2022137103}.

The idea is to apply two consecutive Schur decompositions for the first and third steps of the domain decomposition described above.
\begin{enumerate}

\item At first, we decompose the entire lattice as
\be
	L = \partial\Lambda_0 \cup \left( \bar\Lambda_0 \cup \Lambda_1 \right)
\ee
where $\bar\Lambda_0 = \Lambda_0 \backslash \partial\Lambda_0 = \bigcup_{\hat a} \bar\Lambda_0^{\hat a}$ is the union of the bulk of each local block. Since the two domains $\bar\Lambda_0$ and $\Lambda_1$ are disconnected, it follows that
\be
	\det D = \det D_{\bar\Lambda_0} \det D_{\Lambda_1} \det \tilde D_{\partial\Lambda_0}
\ee
where
\be \label{eq:Dtilde}
	\tilde D_{\partial\Lambda_0} = D_{\partial\Lambda_0} - D_{\partial\Lambda_0,\bar\Lambda_0} D_{\bar\Lambda_0}^{-1} D_{\bar\Lambda_0,\partial\Lambda_0} - D_{\partial\Lambda_0,\Lambda_1} D_{\Lambda_1}^{-1} D_{\Lambda_1,\partial\Lambda_0} 
\ee
is the Schur complement of $D$ with respect to the above decomposition of the lattice. We can notice that the first two terms can already be rewritten as the sum of contributions localized in each local cell only. The last term, instead, is the one which is responsible for global interactions, due to its dependence on the links in the inactive region through $D_{\Lambda_1}^{-1}$.

\item In order to isolate in the last term of Eq.~(\ref{eq:Dtilde}) all the local contributions from links in the same (framed) domain $\Omega_0^{\hat a}$, we can apply a second decomposition, namely $\Lambda_1 = \Phi_1^{\hat a} \cup (\Lambda_1 \backslash \Phi_1^{\hat a})$. Furthermore, for each local domain, we can also separate the contributions coming from its neighboring active cells only, with a procedure that in principle could be iterated to isolate additional, smaller contributions. By following the steps thoroughly described in Section 3 of Ref~\cite{GIUSTI2022137103}, we end up with
\be
        \tilde D_{\partial\Lambda_0} = \hat D_{\partial\Lambda_0} - \hat D_{\partial\Lambda_0,\partial\Pi_1} \hat D_{\partial\Pi_1}^{-1} \hat D_{\partial\Pi_1,\partial\Lambda_0},
\ee
where
\be
	\hat D_{\partial\Pi_1} = D_{\partial\Pi_1} - D_{\partial\Pi_1,\bar\Lambda_1} D_{\Lambda_1}^{-1} D_{\bar\Lambda_1,\partial\Pi_1}
\ee
is the Schur complement of $D_{\Lambda_1}$ with respect to the decomposition $\Lambda_1 = \partial\Pi_1 \cup \bar\Lambda_1$, while
\be
        \hat D_{\partial\Lambda_0} = \hat D_{\partial\Lambda_0}^d + \hat D_{\partial\Lambda_0}^h, \quad \hat D_{\partial\Lambda_0}^d = \sum_{\hat a} \hat D_{\partial\Lambda_0^{\hat a}}, \quad \hat D_{\partial\Lambda_0}^h = \sum_{\hat a \neq \hat a'} \hat D_{\partial\Lambda_0^{\hat a},\partial\Lambda_0^{\hat a'}}
\ee
is the sum of two distinct contributions. The diagonal part $\hat D_{\partial\Lambda_0}^d$ only takes into account interactions that are localized within each cell
\begin{gather*}
        \hat D_{\partial\Lambda_0^{\hat a}} = D_{\partial\Lambda_0^{\hat a}} - D_{\partial\Lambda_0^{\hat a},\bar\Lambda_0^{\hat a}} D_{\bar\Lambda_0^{\hat a}}^{-1} D_{\bar\Lambda_0^{\hat a},\partial\Lambda_0^{\hat a}} - D_{\partial\Lambda_0^{\hat a},\Phi_1^{\hat a}} D_{\Phi_1^{\hat a}}^{-1} D_{\Phi_1^{\hat a},\partial\Lambda_0^{\hat a}} \, ,
\end{gather*}
while the hopping term $\hat D_{\partial\Lambda_0}^h$ contains interactions between two neighboring cells
\begin{gather*}
        \hat D_{\partial\Lambda_0^{\hat a},\partial\Lambda_0^{\hat a'}} = -\frac{1}{2} D_{\partial\Lambda_0^{\hat a},\Phi_1^{\hat a}} \left[ D_{\Phi_1^{\hat a}}^{-1} - D_{\Phi_1^{\hat a}}^{-1} D_{\Phi_1^{\hat a},\partial\bar\Omega_0^{\hat a \ast}} D_{\Phi_1^{\hat a'}}^{-1} \right. \\ \left. + D_{\Phi_1^{\hat a'}}^{-1} - D_{\Phi_1^{\hat a}}^{-1} D_{\partial\bar\Omega_0^{\hat a' \ast},\Phi_1^{\hat a'}} D_{\Phi_1^{\hat a'}}^{-1} \right] D_{\Phi_1^{\hat a'},\partial\Lambda_0^{\hat a'}} \, .
\end{gather*}

\end{enumerate}

The above steps allow to derive the main result we want to discuss, i.e. the block-factorization of the fermion determinant as
\be \label{eq:detD}
	\det D = \frac{\det W_1}{\det D_{\Lambda_1}^{-1} \prod_{\hat a} \left[ \det D_{\Phi_1^{\hat a}} \det D_{\Omega_0^{\hat a}}^{-1}  \right]}
\ee
that leads directly to
\be
	S_F^{\text{eff}}[U] = -\sum_{\hat a} \ln\det D_{\Omega_0^{\hat a}} + \sum_{\hat a} \ln\det D_{\Phi_1^{\hat a}} - \ln\det D_{\Lambda_1} - \ln\det W_1 \, .
\ee
The dependence of $\det D$ on the gauge-field configuration in the active region is fully factorized if we only focus on the denominator, since each factor can be independently computed. Though, a small but (still) not negligible global contribution comes from the numerator of Eq.(~\ref{eq:detD}), which is written in terms of the matrix
\be \label{eq:Wz}
	W_z = \left( \begin{array}{c | c} 
z \mathbb P_{\partial\Lambda_{0}} + [{\hat D}^{\rm d}_{\partial\Lambda_{0}}]^{-1} \hat D^{\rm h}_{\partial\Lambda_{0}} &
 W_{\partial\Lambda_{0},\partial \Pi_1} \\[0.25cm] 
\hline\\[-0.325cm]
W_{\partial \Pi_1,\partial\Lambda_{0}} & z \mathbb P_{\partial\Pi_{1}}
\end{array} \right) \, .
\ee
The off-diagonal blocks
\begin{gather} \label{eq:offdiagW1}
W_{\partial\Lambda_0,\partial\Pi_1} = [{\hat D}^{\rm d}_{\partial\Lambda_{0}}]^{-1} \hat D_{\partial\Lambda_0,\partial\Pi_1} = \sum_{\hat a} \mathbb{P}_{\partial\Lambda_0^{\hat a}} D_{\Omega_0^{\hat a}}^{-1} D_{\Phi_1^{\hat a},\partial\bar\Omega_0^{\hat a \ast}}, \\ W_{\partial\Pi_1,\partial\Lambda_0} = \hat D_{\partial\Pi_1}^{-1} \hat D_{\partial\Pi_1,\partial\Lambda_0}
\end{gather}
can be interpreted as hopping terms between the domains $\partial\Lambda_0$ and $\partial\Pi_1$, while $\mathbb P_{\partial\Lambda_{0}}$, $\mathbb P_{\partial\Pi_{1}}$ are projectors acting as the identity matrix on the domains $\partial\Lambda_0$ and $\partial\Pi_1$ respectively. The upper-left block contains an additional term, $[{\hat D}^{\rm d}_{\partial\Lambda_{0}}]^{-1} \hat D^{\rm h}_{\partial\Lambda_{0}}$, which takes into account residual interactions between different blocks in the active region. In order to accomplish our initial aim, we still need to further factorize $\det W_1$.

\begin{figure}[t!]
\begin{center}
\includegraphics[width=0.8\columnwidth]{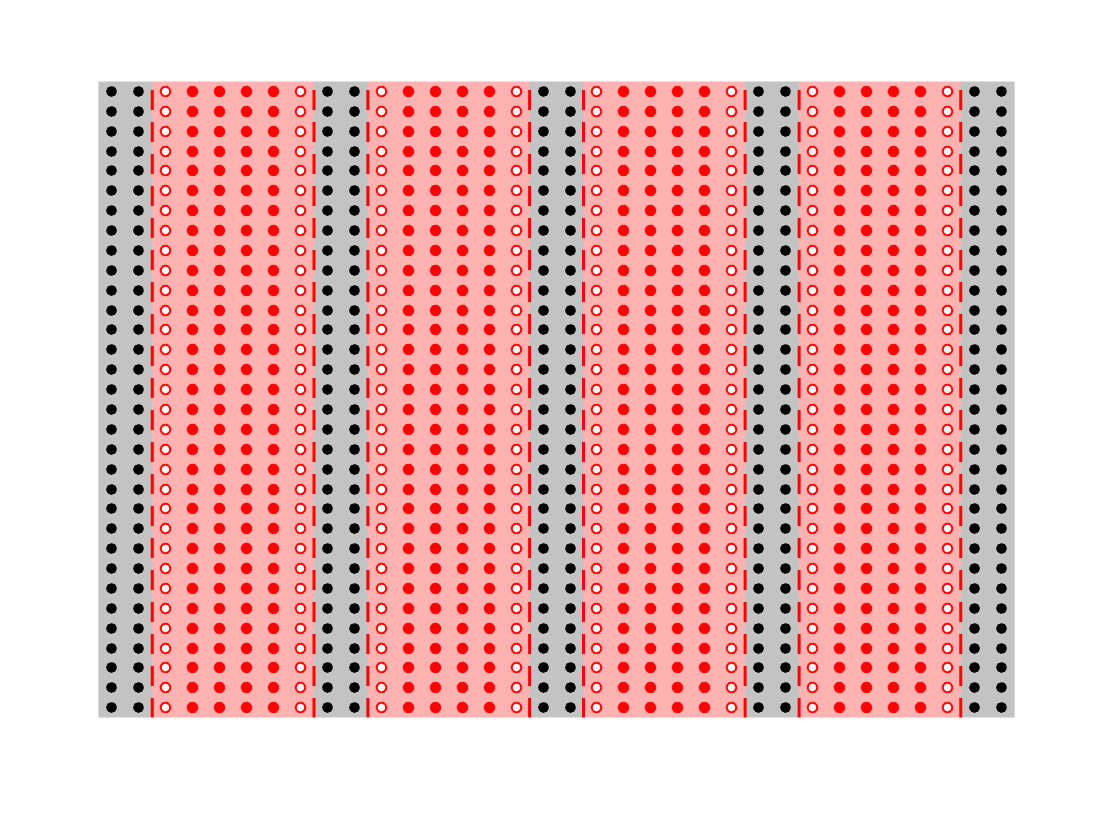}
\caption{One-dimensional domain decomposition of the lattice in thick time-slices in the disconnected domains $\Lambda_{0}$ (red slices) and $\Lambda_{1}$ (grey slices). The empty circles indicate the domain of hyperplanes $\partial \Pi$, that in this case coincides with the internal boundary $\partial\Lambda_{0}$ of $\Lambda_0$.
  \label{Fig:1dDD}}
\end{center}
\end{figure}
Before doing it, we would like to make more sense of this result by showing how it is a (not-so-trivial) generalization of the one-dimensional case that was first discussed in Ref.~\cite{Ce:2016ajy}. We can apply a one-dimensional domain decomposition, as shown in Fig.~\ref{Fig:1dDD}, and adopt the same exact strategy that we described in Section~\ref{sec:2}.
\begin{enumerate}
\item First, we identify the local domains $\Lambda_0^{\hat a}$ as thick time-slices and their union as the disconnected active region $\Lambda_0 = \bigcup_{\hat a} \Lambda_0^{\hat a}$. The rest of the lattice forms the inactive domain $\Lambda_1$, which is a disconnected union of thick time-slices in the one-dimensional case.
\item We can now define the internal boundaries of each block and their union respectively as $\partial\Lambda_0^{\hat a}$ and $\partial\Lambda_0$. When we try to connect the internal boundaries of different slices, we notice that we do not need to include any additional lattice site, i.e. in this case $\partial\Pi_1 = \partial\Lambda_0$ and the domain $\partial\Pi_1$ is actualy empty. Thus, the only block of $W_z$ that is present in this case is the upper-left one, connecting different local slices $\Lambda_0^{\hat a}$.
\item We can finally define the frame $\Phi_1^{\hat a}$ for each local slice and the respective local (framed) slice $\Omega_0^{\hat a} = \Lambda_0^{\hat a} \cup \Phi_1^{\hat a}$. We notice that the union of all the frames takes exactly twice into account each thick time-slice of the domain $\Lambda_1$. 
\end{enumerate}
We can now apply an even-odd decomposition of the domain $\Lambda_0^{\hat a}$, i.e. consider $\hat a \in \{e,o\}$. If we notice that $\det D_{\Phi_1^e} = \det D_{\Phi_1^o} = \det D_{\Lambda_1}$ and, by following the notation of Ref.~\cite{Ce:2016ajy}, we also rename $\Lambda_0^e \rightarrow \Lambda_0$, $\Lambda_0^o \rightarrow \Lambda_2$, we are finally able to exactly recover the one-dimensional result
\be
	\det D = \frac{ \det W_1 }{\det D_{\Lambda_1} \det D_{\Omega_0}^{-1} \det D_{\Omega_2}^{-1} } \, .
\ee
We want to point out that the four-dimensional generalization of this result is not straightforward due to the presence of two additional complications:
\begin{enumerate}
\item the existence of corners, i.e. the domain $\partial\Pi_1$ and the off-diagonal terms of $W_1$;
\item $\Lambda_1$ being a connected domain.
\end{enumerate}
In order to accomplish our initial aim of factorizing $\det D$, we need to address the remaining issue of factorizing $\det W_1$.

%% file: sec4.tex
In order to factorize the dependence of $\det W_1$ on the gauge-field configuration within the active region $\Lambda_0$, we need to further analyze the structure of the matrix $W_1$ by making two observations.
\begin{enumerate}
\item The first observation to be made is that this matrix can be rewritten as the sum of the identity matrix on the domain $\partial\Pi$ and off-diagonal terms which are suppressed with the thickness of $\Lambda_1$. More in particular, the term $[{\hat D}^{\rm d}_{\partial\Lambda_{0}}]^{-1} \hat D^{\rm h}_{\partial\Lambda_{0}}$ connects two different blocks in the active region, while the off-diagonal blocks of $W_1$ defined in Eq.~(\ref{eq:offdiagW1}) propagate from one local cell in the active region to the inactive one or viceversa. Therefore, the off-diagonal terms of $W_1$ are all suppressed at least with the thickness of the inactive region $\Lambda_1$.
\item The only dependence of $W_1$ on the links in the active region is in the off-diagonal terms in the first line.
\end{enumerate}
The general idea is that, if $\Lambda_1$ is thick enough, namely at least $0.5$ fm \cite{DallaBrida:2020cik}, then the off-diagonal terms of $W_1$ are tiny. In this case, a polynomial approximation of $\det W_1^{-1}$ of order $N$ is expected to be reliable at a small value of $N$, namely of order $10$ or so. First, we need to define the approximating polynomial of $1/z$ as
\be
	P_N (z) = \frac{1-R_{N+1}(z)}{z} = c_N \prod_{k=1}^N (z-z_k)
\ee
such that
\be \label{eq:WW}
	\frac{1}{\det P_N(W_1)} = C \prod_{k=1}^{N/2} \det \{ (z_k - W_1)^\dagger (z_k - W_1) \}^{-1} = C \prod_{k=1}^{N/2} \det (W_{u_k}^\dagger W_{u_k})^{-1},
\ee
\be
	u_k = 1-z_k = \cos{ \left( \frac{2\pi k}{N+1} \right) } + i\sqrt{1-c^2} \sin{\left( \frac{2\pi k}{N+1} \right) }, \, k=1,\dots,N \, .
\ee
We can now introduce $N/2$ multi-boson fields~\cite{Luscher:1993xx,Borici:1995np,Borici:1995bk,Jegerlehner:1995wb} that reside on the global, connected domain $\partial\Pi$ to represent each term in the product of Eq.~(\ref{eq:WW}) as
\be
	\frac{1}{\det (W_{u_k}^\dagger W_{u_k})} \propto \int d\chi_k d\chi_k^\dagger e^{-|W_{u_k}\chi_k|^2}
\ee
with the multi-boson action defined as
\be \label{eq:MBaction}
	S_{\text{MB}}[\chi,\chi^\dagger,U] = |W_{z} \chi|^2 = \sum_{\hat a} S_{\text{MB}}^{\hat a}[\chi,\chi^\dagger,U_{\Omega_0^{\hat a}}] + S^r_{\text{MB}}[\chi,\chi^\dagger,U] \, ,
\ee
where
\be \label{eq:localMB}
	S_{\text{MB}}^{\hat a}[\chi,\chi^\dagger,U_{\Omega_0^{\hat a}}] = \Big|P_{\partial\Lambda_{0}^{\hat a}}\Big[z\,\chi_{_{\partial\Lambda_{0}}} +
{\hat D}^{-1}_{\partial\Lambda_{0}^{\hat a}} \hat D^{\rm h}_{\partial\Lambda_{0}} \chi_{_{\partial\Lambda_{0}}}
+ D_{\Omega^{\hat a}_{0}}^{-1}\, D_{\Phi^{\hat a}_{1},\partial \bar\Omega^{\hat a *}_0} \chi_{_{\partial\Pi_1}}\Big] \Big|^2
\ee
and
\be \label{eq:globalMB}
	S^r_{\text{MB}}[\chi,\chi^\dagger,U] = \Big|z\, \chi_{_{\partial\Pi_1}} + W_{\partial \Pi_1,\partial\Lambda_{0}} \chi_{_{\partial\Lambda_{0}}} \Big|^2 \, .
\ee
The second observation we made above now proves to be crucial: the only dependence of the multi-boson action in Eq.~(\ref{eq:MBaction}) on the gauge field in the active region is in the first term, thus making it fully factorized.

We can also define a reweighting factor
\be \label{eq:rw}
	\mathcal W_N = \det \{ 1-R_{N+1}(W_1) \}
\ee
so that
\be \label{eq:detDN}
	\frac{\det D}{\mathcal W_N} \propto \frac{1}{\det D_{\Lambda_1} \prod_{\hat a} \left[ \det D_{\Phi_1^{\hat a}} \det D_{\Omega_0^{\hat a}} \right] \prod_{k=1}^{N/2} \det (W_{u_k}^\dagger W_{u_k}) } \, .
\ee
This allows us to compute averages by means of a reweighting procedure as
\be
	\langle O \rangle = \frac{ \langle O \mathcal W_N \rangle_N }{\langle \mathcal W_N \rangle_N}
\ee
where the averages $\langle \cdot \rangle_N$ are computed with the factorized fermionic effective action
\be \label{eq:SFeffN}
\begin{gathered}
	S_F^{\text{eff},N}[U,\{\chi_k,\chi_k^\dagger\}_{k=1}^{N/2}] = \sum_{\hat a} \left\{ -\ln\det D_{\Omega_0^{\hat a}} + \sum_{k=1}^{N/2} S_{\text{MB}}^{\hat a}[\chi_k,\chi_k^\dagger,U_{\Omega_0^{\hat a}}] \right\} \\
	+ \sum_{\hat a} \ln\det D_{\Phi_1^{\hat a}} - \ln\det D_{\Lambda_1} + \sum_{k=1}^{N/2} S^r_{\text{MB}}[\chi_k,\chi_k^\dagger,U] \, .
\end{gathered}
\ee
Our initial goal has been achieved: the dependence of the fermion determinant on the gauge field configuration in the active region has been fully factorized, see Eq.~(\ref{eq:detDN}), up to the small reweighting term in Eq.~(\ref{eq:rw}) that can be included in the observable. This in turn implies that the fermionic effective action can be rewritten as in Eq.~(\ref{eq:SFeffN}). On one hand, each term in the sum, see Eq.~(\ref{eq:localMB}), only depends on the gauge field configuration in the (framed) local domain $\Omega_0^{\hat a}$ and therefore it can be computed independently from the others. On the other hand, all the global contributions only come from the second term, see Eq.~(\ref{eq:globalMB}), which only depends on the gauge field configuration in the inactive region, and therefore it is constant if we only update links in the active domains.

%% file: concl.tex
We have achieved our initial goal of factorizing the dependence of the fermion determinant on the gauge field configuration in the active region. We managed to do so by means of a judiciously chosen four-dimensional overlapping domain decomposition of the lattice that led to a block-factorization of the fermion determinant. The small remaining global term was treated by means of a multi-boson representation, where the multi-boson action is factorized as well. The only residual factor can be regarded as a reweighting term to be included in the observable.

Besides showing this thoretically appealing factorization, we would like to conclude by describing some of its possible applications.

{\it Multi-level integration $-$} This factorization can be immediatly employed for a multi-level integration scheme, where we alternate updates of the whole lattice with local updates of the active region at fixed gauge field configuration in the inactive domain. We need to ensure that a good fraction of the link variables can be updated in each step. For instance, we can consider blocks with an extension of $2.5$ fm and a frame of $0.5$ fm in all directions so that around $50\%$ of the links are kept active. Extensive numerical tests for the one-dimensional domain decomposition~\cite{Ce:2016idq,Ce:2016ajy,Giusti:2017ksp,Ce:2017ndt,DallaBrida:2020cik} have already shown the benefit of the multi-level integration in solving the signal to noise ratio problem, which was first identified in~\cite{Parisi:1983ae}, in the computation of correlation functions in lattice QCD.

{\it Master-field $-$} With this factorization, during the molecular dynamics evolution and for the accept-reject step, an inversion of the global lattice Dirac operator is never required. The only global inversion needed is for multi-boson fields generation, but it is applied on a vector belonging to the domain $\partial\Pi$ which is much smaller than the entire lattice. In master field simulations in the presence of fermions~\cite{Luscher:2017cjh,Giusti:2018cmp,Francis:2019muy}, this mitigates the problem of the increasing numerical precision needed for inverting the Dirac operator on larger and larger volumes.

{\it Flow-based generative models $-$} In the context of flow-based Markov Chain Monte Carlo methods~\cite{Albergo_2019,Kanwar_2020,Abbott_2022}, rather than approximating the full distribution, one can imagine to employ the local approximation in Eq.~(\ref{eq:SFeffN}) as the desired distribution. The network will then be trained to approximate it and the self-training and accept-reject (or reweighting) steps usually applied in this context will be combined with the reweighting procedure described in this paper.

{\it Parallelization $-$} The factorization explicitly decouples the link variables in different blocks from each other, allowing the HMC molecular-dynamics evolution and accept-reject steps to be run independently in each block. On heterogeneous architectures, one can envisage to simulate each block on a sub-set of nodes which have faster connections (or, for instance, on a single GPU) without the need to communicate during long periods of simulation time. A communication overhead is required only for global updates of the gauge field configuration on the whole lattice and for the generation of the multiboson fields. This is typically a very small fraction of the computer time of the simulation.